**Vol. 2. No.1**
**"Pedagogy, Education and Innovation in 3-D Virtual Worlds"**
**April 2009**

**Guest Editors**
Leslie Jarmon
Kenneth Y. T. Lim
B. Stephen Carpenter

**Editor**
Jeremiah Spence

**Technical Staff**
Andrea Muñoz
Amy Reed
Barbara Broman
John Tindel
Kelly Jensen

# Second Life physics:
## *Virtual, real or surreal?*

By Renato P. dos Santos, ULBRA - Universidade Luterana Brasileira, Brasil

## Abstract

*Science teaching detached itself from reality and became restricted to the classrooms and textbooks with their overreliance on standardized and repetitive exercises, while students keep their own alternative conceptions. Papert, displeased with this inefficient learning process as early as 1980, championed physics microworlds, where students could experience a variety of laws of motion, from Aristotle to Newton and Einstein or even "new" laws invented by the students themselves. While often mistakenly seen as a game, Second Life (SL), the online 3-D virtual world hosted by Linden Lab, imposes essentially no rules on the residents beyond reasonable restrictions on improper behavior and the physical rules that guarantee its similitude to the real world. As a consequence, SL qualifies itself as an environment for personal discovery and exploration as proposed by constructivist theories. The physical laws are implemented through the well-known physics engine Havok, whose design aims to provide game-players a consistent, "realistic" environment. The Havok User Guide (2008) explicitly encourages developers to use several tricks to cheat the simulator in order to make games funnier or easier to play. As it is shown in this study, SL physics is unexpectedly neither the Galilean/Newtonian "idealized" physics nor a real world physics virtualization, intentionally diverging from reality in such a way that it could be called hyper-real. As a matter of fact, if some of its features make objects behave "more realistically than real" ones, certain quantities like energy have a totally different meaning in SL as compared to physics. Far from considering it as a problem, however, the author argues that its hyper-reality may be a golden teaching opportunity, allowing surreal physics simulations and epistemologically rich classroom discussions around the "what is a physical law?" issue, in accordance with Papert's never-implemented proposal.*

# Second Life physics:
## *Virtual, real or surreal?*

By Renato P. dos Santos, ULBRA - Universidade Luterana Brasileira, Brasil

Among all the virtual worlds that exist today, Second Life (SL), which appeared in 2004 almost out of the blue, has now the largest user base. Its older competitors like AlphaWorld, Active Worlds, the Croquet Project, and other virtual worlds, all have different strengths, but none of them matches the popularity with the general public and the commercial companies that SL has conquered (Bestebreurtje, 2007).

SL is an online 3-D virtual community developed by Linden Research, Inc. founded by Philip Rosedale, better known in the SL world as *Philip Linden* (*Linden Lab Management*, n.d.). SL is hosted and operated by Lab of Linden Research, Inc., also known as *Linden Lab*. The entire world of SL, called *the grid*, (including all avatar data, objects, landscapes, textures, and texts) is hosted on servers run by Linden Lab. SL is still seen as a game, but its residents have disputed this notion because there are essentially no rules imposed on the residents (Bestebreurtje, 2007). The only exceptions are the restrictions on areas that are not open to the general public and the physical rules that make objects to interact realistically. To access it, users only have to download and install client software locally. Once logged, SL users, called *residents*, can walk around, explore the world, enjoy the 3-D scenery, fly, drive cars and other vehicles, interact with other avatars, play, or create objects. There are a wealth of resources for building complex objects, with many different textures, such as chairs, clothes, jewels, vehicles, guns, and even entire buildings. In fact, most of SL world has been built by the residents themselves, which has been characterized as a shift of culture from a media consumer culture to a participatory culture (Jenkins, 2006).

The ease in which new users can join SL, combined with support from several educational and library groups, discussion forums and a wide range of free communication, graphics, design, and animation tools, makes many educators from around the world see SL as a versatile environment to conduct pedagogical activities (Calogne & Hiles, 2007). Bradley (2008), for example, relates how in his Introduction to Organic Chemistry students created a life-size model of a molecule around which the teacher is able to walk, with the students, and comment real time. Calogne and Hiles as well as Conklin (2007) list various educational uses of Second Life, including Art, Law, Religion, English, Programming, Geography, Politics, Economy, Mathematics, Biology, and physics teaching.

The first idea that comes to mind is to use SL to offer courses online. But, this author aligns himself with Eliëns, Feldberg, Konijn & Compter (2007) in considering this approach rather naïve and outdated while there are other much more appealing alternatives such as simulations and modeling  (Borba & Villarreal, 2005).

Although many authors stress the SL "potential" for simulations that promote physics learning, this author did not succeeded in finding any concrete example of physics simulation. There are, however, many artifacts left to free manipulation by visitors in places like the *Institute of Physics Experimenta*, situated in the coordinates *Rakshasa (207, 26, 25)*. Furthermore, researchers at Denver University are planning to build the first virtual nuclear reactor to train new nuclear engineers (Medeiros, 2008).

The physical verisimilitude of the metaverse, in the sense that an avatar cannot pass through walls and stones tossed into water will behave as expected, relies on the obedience to physical laws and principles such as gravity, buoyancy, mass, friction, and so forth. This obedience is usually ensured by means of third-party software called *physics engine*. Vehicles are one application of the physics engine.

Differently from other metaverses, such as *Google Lively*, where physical laws are not seriously taken into account, SL is possibly the most realistic virtual environment in the market, as objects are controlled by the Havok software. This powerful software has been used in creating many internationally acclaimed films over the years such as *Troy*, *X-Men: The Last Stand*, *Harry Potter and the Order of the Phoenix* and *The Chronicles of Narnia: Prince Caspian*, among others (Havok in the movies, n.d.).

In this work, SL physics will be studied through a comparative analysis between the Newtonian physics taught in schools and the SL world physical features and physically interesting LSL functions, as described in the following sources, in addition to the author's own experiences:

- *Havok Physics Animation v. 6.0.0 PC XS User Guide* ;
- *LSL Portal* ;
- Guidelines for educators;
- *LSL Wiki;*
- *SL Wiki;* and
- *SL Wikia*.

Also discussed are the differences found between SL physics and that physics that is taught at School within the framework of the notions of *Reality in Science* and of *Virtuality*, according to Lévy (1998) and Eco (1986).

After this analysis, pedagogical implications and alternatives, based on Papert's (1993) never-implemented *physics microworlds* proposal will be discussed.

## Second Life Physics Analysis

*"Morpheus: [. . .] yet their strength and their speed are still based on a world that is built on rules" (Irwin, 2002).*

The Second Life world consists of many interconnected, uniquely named simulators, referred to as *sims* or *region*s. Each simulator keeps track of the objects and agents within its region, simulates physics, runs scripts, and caches and delivers object and texture data within the sim to clients.

The aim in Havok's (2008) design is to provide simulation that gives the game-player a consistent environment to explore (p. 374). In principle, Havok deals with Newtonian mechanics, or the high school laws of motion that describe the behavior of objects under the influence of other objects and external forces (p. 375).

SL, through its Linden Scripting Language (LSL), offers resources to attach behaviors to objects such as fountains, guns, or vehicles so that an object can change its color, size, or shape, while it can move, listen to your words, talk back to you, or even talk to other objects. LSL follows the familiar syntax of a C/Java-style language and features almost 400 built-in functions for manipulating physics and avatar interaction, many of which of special interest to physical studies in this metaverse. For example, *llGetPos()*[2] and *llGetVel()* return vectors that are the object's region position and velocity, respectively; *llGetOmega()* returns its angular velocity, while *llGetForce()* and *llGetTorque()* return vectors representing the force and the torque, respectively, acting upon the object.

As Havok deals with game genre specific problems like vehicle simulation, human ragdolls, physical interaction of keyframed characters within a game environment, and character control (2008), it does not even try to simulate any physics beyond Mechanics, excluding any possible electromagnetic or nuclear interactions.

While many physical quantities have their physical counterpart in SL, certain quantities have quite different definitions in SL when compared to the Newtonian physics ones, as will be seen below.

*Time*

Still today, time, one of the most fundamental physical quantities, usually refers to the classical Newtonian conception of an "absolute" and "equably" flowing time, used to compare the intervals between events and their durations and to sequence them, therefore making possible to quantify the motions of objects and to formulate a prescription for the synchronization of clocks. This, of course, is quite different from Einstein's proposal of a new method of synchronizing clocks using the constant, finite speed of light as the maximum signal velocity, which gave birth to the Theory of Relativity.

In SL, being a Massive Multiplayer Online Reality Game, time is needed to keep things moving in (or out of) sync with everything else. However, SL physics can be impacted by network lag and server load, and therefore may not be particularly accurate. All physics and scripts generate simulator lag which can make avatars experience a slowed-down (slow-motion, "bullet-time") movement from its usual region frames per second (FPS) value of 45.0, as returned by the *llGetRegionFPS( )* function. When the sim server cannot keep up with the processing of its tasks, it will use a method called *time dilation* to cope with it. Time dilation will slow script time and execution down to the limit when time dilation value reaches zero and script execution halts. The function *llGetRegionTimeDilation()* returns the current region simulator time dilation, the ratio between the change of script time to that of *real world* time, as a float value that ranges from 0.0 –full dilation – to 1.0 – no dilation (*LSL Wiki*, llGetRegionTimeDilation and *LSL Portal*, llGetRegionTimeDilation). A collection of lag reduction tips is provided at *LSL Wiki* under *Lag*.

While SL is able to run qualitative experiments and to cope with simple mechanics experiments with a corresponding decrease in accuracy, it will definitely not give response time down to milliseconds consistently (*Guidelines for educators,* 2008, Technical essentials, § 5).

*Mass*

The concept of mass is one of those basic physical concepts whose real significance is never fully disclosed in textbooks or lecture courses (Jammer, 1997). In fact, it is a rather abstract concept whose meaning evolved from the ancient metaphysical opposition of matter and spirit to the present relativistic equivalence to energy or to space-time curvature. However, its most common meaning is still related to the object's resistance to accelerate when a force is applied on it.

In SL, mass is the measure of translational inertia, the tendency of a body to resist accelerations, expressed in lindograms (Lg). The mass of an object is reported by functions *llGetMass()* and *llGetObjectMass()*. For example, a typical avatar mass is around 2 Lg.

Contrary to the Newtonian (1947) definition of mass as arising from the product of its an object's density and volume, in SL object mass depends only on its size and shape, not on its material type, set through the constant *PRIM_MATERIAL* to one of the eight different available materials, such as glass, metal, flesh, and so forth. However, avatar mass depends only on its height, irrespective to its fatness, thickness, muscle, or other factors. Attachments will not alter avatar mass, except for shoes, which change avatar height and therefore its mass.

*Gravity*

In physics, gravity – or, more generally, gravitation – refers to the natural phenomenon by which objects attract one another. A direct consequence of it is the well-known weight force that every object experiences.

In SL, every physics-enabled object with mass $m$ will be subject to a constant force $P$ given by

$$P = m \cdot 9.8 \, \text{m/s}^2$$

applied in the negative z-direction to simulate the acceleration under gravity (*LSL Wiki*, n. d., Gravity). However, the function *llSetBuoyancy()* can cancel the effects of gravity, as seen below.

*Acceleration*

In Havok (2008), the quantity acceleration has the usual physical definition as "the rate of change of velocity over time" (p. 380). In Newtonian terms, the acceleration $a$ of an object with mass $m$, when subject to the action of a force $F$, is given by

$$a = F/m.$$

However, the SL function *llGetAccel()*, instead of returning its acceleration, returns the vector

$$(\text{llGetForce}()/\text{llGetMass}()) + <0,\ 0,\ -9.8>$$

which is a sum where the first term corresponds to the dynamical object acceleration, due to the action of some force, as seen above, and the second term is a vertical downwards acceleration that simulates the terrestrial gravity effect (*LSL Wiki*, n. d., llGetAccel).

For an object resting on the floor, the value for its acceleration, as returned by *llGetAccel()*, should be always <0, 0, 0>, which corresponds to real world physics, if we consider a normal force acting on the object by the floor which exactly offsets the force of gravity.

However, random, fast-changing values for *llGetAccel()* were observed for a still object, as shown in Figure 1. This is possibly due to the action of the omnipresent SL "wind," whose velocity can be obtained through the function *llWind()* (*LSL Wiki*, n. d., llGetAccel),.

Although aerodynamic or hydrodynamic effects of viscosity with air or water have not been implemented, the value for the acceleration of a free-falling object, initially equal to the gravitational acceleration, will be gradually reduced to zero in order to *simulate* the air resistance effect that makes the objects asymptotically approach a *terminal velocity*, as in the real world. On the other hand, as there is no "real" air resistance, "any impulse off the vertical (gravity) axis will cause the object to keep moving forever" (*LSL Wiki*, n. d., llApplyImpulse).

As a consequence of the points above, users are advised to trackvelocity and measure its changes to get the actual acceleration values, which means to use the well-known acceleration formula

$$a = \frac{\Delta v}{\Delta t},$$

where $\Delta v$ denotes a velocity change during a certain time interval $\Delta t$.

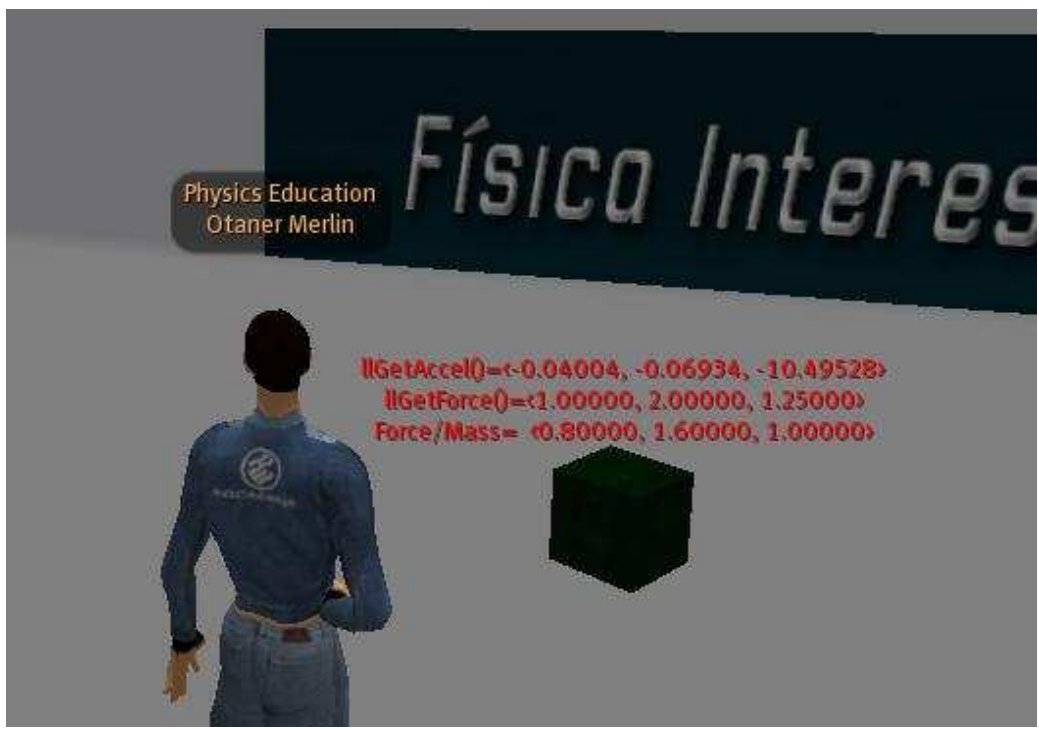

**Figure 1. Random Acceleration in a Still Object.**

*Energy*

In physics, the concept of energy was developed by various scientists since the eighteenth century, initially with the purpose of studying free falling objects and collisions, and it was formalized through the discovery of various conversion processes until the middle of the nineteenth century. It is a very abstract, theoretical concept and its conceptualization comes from the principle of conservation of energy (Elkana, 1974). Throughout history, various forms of energy were defined, including kinetic, potential, electric, and chemical, which are interconvertible.

In Havok (2008), "energy management is concerned with identifying objects in a scene that are not doing very much and removing these from the physical simulation (known as *deactivating* or turning off the object) until such time as they begin to move again" (p. 379). Therefore, SL "energy" is somehow nearer to *activity* in Havok.

In SL, energy is a dimensionless quantity ranging between 0.0 and 1.0 and used to control how effectively scripts can change the motion of physical objects: if the energy of an object is 100 percent, LSL functions will have their full effect; yet, if it has a value lower than 100 percent, the same action on the same object will have a proportionally smaller effect, and when it reaches 0, actions will have no effect at all (*LSL Wiki*, n. d., Energy).

Instead of the usual physics kinetic energy, calculated in terms of its mass $m$ and its velocity $v$ as $E = \frac{1}{2}mv^2$, in SL, energy behaves more like a momentum transfer $p$ ($p = mv$). This makes it easier to achieve high speeds with light objects (small *m*) but vehicles will probably cause smaller damages in collisions.

An object expends energy when scripts call functions to change its motion but objects continuously receive energy from the SL grid through a stipend at a rate of 200/mass units of energy per second until the 1.0 full energy limit. Kinetic functions demand energy at different rates and some of them may even not be able to act on heavy objects if they "eat" energy faster than the grid can refill it (*LSL Wiki*, n. d., Energy). See, for example, the function *llSetBuoyancy()* below.

Contrary to physics, where energy is defined only for the system where the object is inserted, in SL, energy is stored inside the object, in an impulse energy "reservoir" proportional to its mass (*SL Wikia*, n. d., Impulse Energy). Each time an object is rezzed (the act of making an object appear in this metaverse by dragging it from a resident's inventory or by creating it via the edit window) or its mass is changed (as in by changing its size or shape), its energy is reset to 0.

It should be noted, however, that the "real" object energy value is inaccessible, since the *llGetEnergy()* function merely returns object activity "as a percentage of maximum" (*LSL Portal*, n.d., llGetEnergy()). But, contrary to what one could think, in the real world, too, one cannot directly access the energy of a system by a simple measurement – there is no energy-meter. As Sexl (1981) explains, "the total energy of a system has to be *calculated* from observable quantities like velocities, distances, charges and so on." This is the exact procedure one will have to follow to obtain the energy value in SL.

*Friction*

In physics, friction is the contact force resisting the relative lateral motion of solid and/or fluid surfaces in contact. It depends on the normal force exerted between the surfaces, as a result from the objects' weights; heavier bodies will show higher friction. It also depends on the materials in contact and on their surface roughness, both quantified by a *coefficient of friction*.

In SL, friction is defined as "the effect of multiple collisions with other objects, e.g. air or water molecules" (*LSL Wiki*, n. d., Friction). This definition, however, describes other effects known as *drag forces*, such as air and water resistance, which also depend on the speed of the object moving through the medium. Moreover, drag forces, which increase with object speed, can reach the force of gravity value, canceling it and imposing a constant *terminal velocity* on the object (see *Acceleration* above).

In SL lighter, less massive objects are simply affected to a lesser extent by friction than heavier, more massive ones under identical conditions. Although there is no coefficient of friction to be altered, an object's friction can be indirectly changed by modifying its material type among the available ones, listed here from least to most friction: glass, metal, plastic, wood, plastic, rubber, stone, and flesh (*LSL Wiki*, n. d., Material). On the other hand, object buoyancy, as set by function *llSetBuoyancy()*, affects friction (*LSL Wiki*, llSetBuoyancy).

As before mentioned, aerodynamical or hydrodynamical friction effects were not implemented in SL, except for air resistance on free falling objects.

*Buoyancy*

In physics, buoyancy refers to the vertical upward force on an object, such as a ship or a balloon, exerted by the surrounding liquid or gas and equals the weight of the fluid displaced by the object, according to Archimedes's principle.

In SL, the buoyancy of an object is a dimensionless quantity and is set by the function *llSetBuoyancy()*. The default value is 0.0. Values between 0.0 and 1.0 mean a gentler than regular fall, the closer to 0.0 the closer to normal behavior. Setting this to exactly 1 will cause the object to float as if no gravity exists and buoyancy greater than 1.0 will make it to rise. Negative buoyancy values are allowed and will simulate a downward force, which will nonetheless also cause the object to vibrate considerably on the ground while the physics engine tries to "settle" it (*LSL Wiki*, n. d., llSetBuoyancy).

Buoyancy applies to "physical" objects only, as seen below. It is often used to make an object like a balloon float up slowly, as if gravity did not affect it. Wind can cause the object to drift.

It must be noticed that water has limited meaning in SL and buoyancy does not take water level into account: the object will float up the same rate whether it is under or above water. Unlike some other characteristics, this is cancelled if the script that sets buoyancy is removed from the object. This function drains energy to keep the object floating. Therefore, it will not make a 90 kg object to hover, unless an extra upward force is applied.

As one sees, SL buoyancy is quite different from the real life one and to get a more accurate value for buoyancy, a simple llSetBuoyancy call does not suffice. It could be calculated from the object's displacement based on volume and on the density of the material being displaced.

*Light*

As in the most primitive conceptions of light (LaRosa, Patrizi, & Vicentini-Missoni, 1984), it is a phenomenon totally pervasive in SL life. It is part of the environment, simply "is there" without any physical mechanism involved in its production or propagation. As expected, the main illumination comes from the sun and the moon replicas in SL, and their direction and their light intensity are uniform not only over the entire simulator but over the entire world. The sun and the moon are directly opposed to each other at all times, and as a result, the moon always appears full. Their orbital centers and speed are such that SL day and night corresponds to three and one RL hours respectively. Private island owners are even able to fix the position of the sun irrespectively of the direction seen in most of the world (*LSL Wiki*, n. d., llGetSunDirection()).

On the other hand, any object, to a limit of six per region, can be made a light source through the mere activation of a checkbox in its properties list, while the color and intensity of the emitted light can also be easily set, as well as the distance it will reach and even its intensity falling-off.

*Physical and "Phantom" Objects*

It must be noted, however, that all that has been mentioned above applies only to objects made *physical* through the constant *STATUS_PHYSICS*. The attribute *physical* essentially enables inertia and gravity to act on the object which can then be moved and rotated using kinetic functions such as *llMoveToTarget()*, *llSetForce(,)* or *llApplyRotationalImpulse()*. A moving physical object will follow the real world rules and come to a halt and settle, but that depends on the object's velocity and/or its buoyancy, as a negatively buoyant object proportionally bounces around more before it settles. An object subject to a continuous impulse/collision will tend to never settle, however, unless it interpenetrates something.

Objects may also be made "phantom" by means of the constant *STATUS_PHANTOM*, in the sense that it can freely pass through anything except the terrain, objects, and avatars, without collision; phantom objects are not transparent, however. Curiously, phantom objects can be made physical when they will start colliding with land; the difference between phantom and non-physical objects is not clear.

**On the Realism of Second Life Physics**

*"Morpheus: What is real? How do you define real? If you're talking about what you feel, taste, smell, or see, then real is simply electrical signals interpreted by your brain"* (Irwin, 2002).

*"Cypher: I know this steak doesn't exist. I know that when I put it in my mouth, the Matrix is telling my brain that it is juicy and delicious"* (Irwin, 2002*).*

Bachelard (1934) argued "every fruitful scientific revolution has forced a profound revision in the categories of the real" (p. 134). Therefore, in order to fully understand the present thesis that, while SL physics is neither the Galilean/Newtonian physics nor a real world physics virtualization but intentionally diverges from reality in such a way that it may be labeled *hyper-real,* SL provides a richer environment for physics teaching than (still) positivistic lectures and classical simulations, a brief historical discussion covering those different worldviews is needed.

*Real world ancient physics*

Dynamics and kinematics appear in antiquity as philosophical discussions only. The Aristotelian "Law" *nullum violentum potest esse perpetuum*[*] implies that as soon as the force applied to the body ends, its motion ends too. This is in good accordance to our everyday experience, to the *intuitive physics* we develop by ourselves by interacting with our environment, and which will later constitute a serious learning obstacle to the Newtonian physics (McCloskey, 1983).

According to Dugas (1955), Aristotle's intuitive theories have their origin in observations most routinely made in daily life, as they take the passive resistances to motion in account. Therefore, Aristotle tried to build a *real world physics*.

*Ideal-world classical physics*

In the Scientific Revolution, Galileo avoided the endless medieval discussions about the Aristotelian causes and focused on describing the course of the motion of falling bodies and projectiles in an exact mathematical manner (Dijksterhuis, 1969). As a matter of fact, he arrived at the correct conclusion that all bodies of any substance fall with the same acceleration only by *abstraction* from the air resistance and considering motion in vacuum, a method which paved the path for the Newtonian perpetual inertial motion (Crombie, p. 298). Doing so, Galileo put himself – and physics – immediately and conscientiously *out of reality* as an absolutely slick plane, an absolutely round sphere, both absolutely rigid, which are not found in our physical reality (Koyré, 1978) but only in Plato's World and, of course, in the abundant, standardized, and repetitive end-of-chapter textbooks exercises. This, the Physics still taught in schools, restricted to the classrooms and positivistic laboratories — hard, abstract and detached from student's everyday reality — is an *ideal-world physics*.

*Counter-intuitive modern physics and surrealism*

The introduction of technologies that enhanced the vision, such as the *camera obscura*, the telescope, the zoopraxiscope, and many others, allowed for the creation of images that were disconnected from the tangible and began to *define* the real (Jones, 2006). A well-known example is Eadweard Muybridge's pioneer study on the horse motion (Figure 2)

Journal of Virtual Worlds Research - Second Life Physics 13

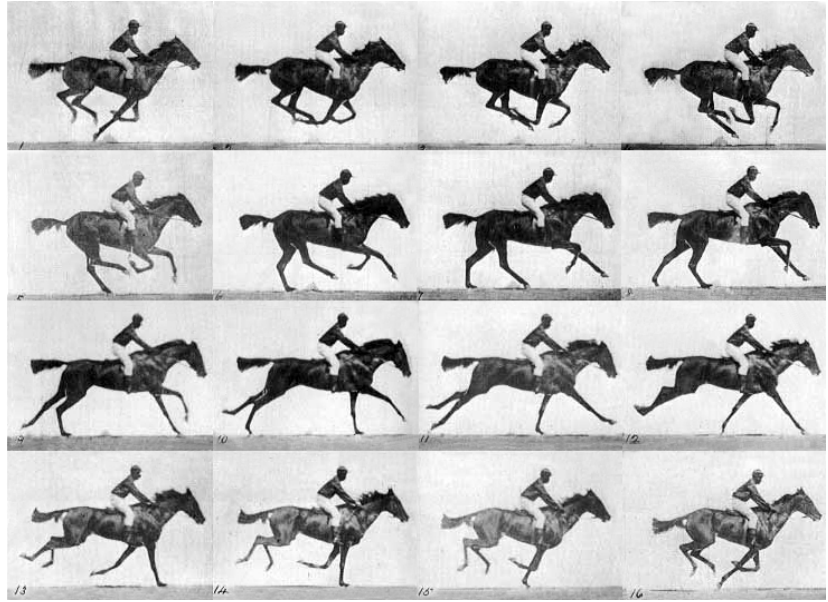

**Figure 2.** *The Horse in Motion*, **Eadweard Muybridge (1878).**

It showed that horses never fully extended its legs forward and back, hooves all leaving the ground (Eadweard Muybridge, 2009), as contemporary illustrators tended to imagine (Figure 3):

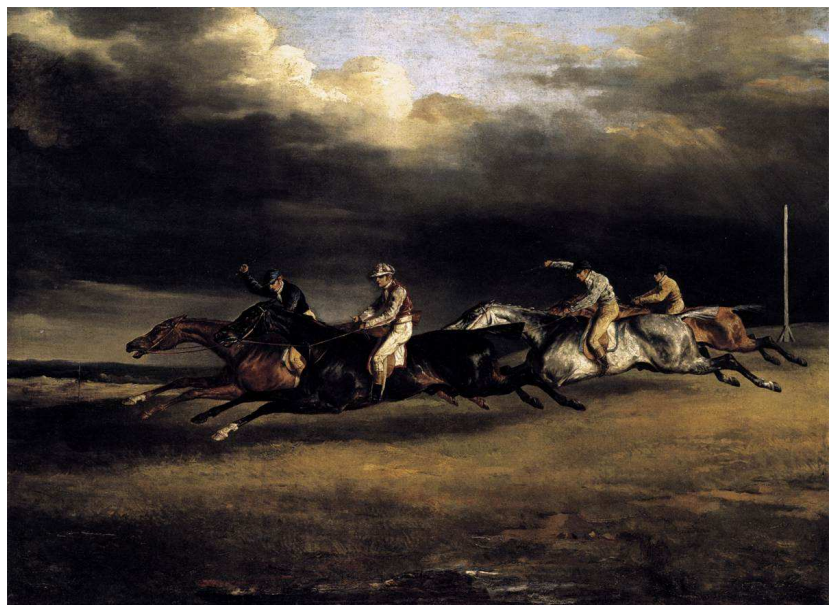

**Figure 3.** *Le Derby d'Epsom*, **Théodore Gericault (1821).**

Such an "obvious" common sense, horse legs position was unfortunately not true. The real position is non-intuitive and intangible, made conceivable only through an instrument: phenomenotechnics (Bachelard, 1934, p. 12).

On the turning of the century, our physical intuition took a major blow from Quantum Mechanics, Relativity, and Chaos Theory with all their concepts such as entanglement, gravitational lensing, and Hausdorff dimension that nobody understands (Feynman, 1967, p. 129). This fascinating physics populates imagination, sci-fi movies, and books but is unfortunately almost absent from classrooms, reserved to physics majors. At the same time that new physics was revolutionizing Classical Science and Popper's Critical Rationalism was refuting the Classical Scientific Method, Modern Art was questioning the axioms of the previous age and struggling to find a language for the "new reality" revealed by the physicists, trying to capture the essence of the schizoid wave-particle duality (Parkinson, 2008). Surrealism in particular, according to Breton (1971), is based on the belief in the superior reality of certain forms of association heretofore neglected. Most popular context of the word "surrealism" is associated to political or social critiques through art but a few surrealist artists went further and criticized the limited rationalistic and positivistic physics worldview through idiosyncratic appropriation and juxtaposition of its parts as a means for imagination to recover its rights (Breton, 1971). While the Copenhagen Interpretation of Quantum Mechanics states that we cannot talk about the real, but only the representations we make of it, Magritte produced various paintings in which he explores and denounces, in a witty and thought-provoking style, the difference between the real world and its representation, such as in his well-known 1935 work *La Condition Humaine* (Figure 4).

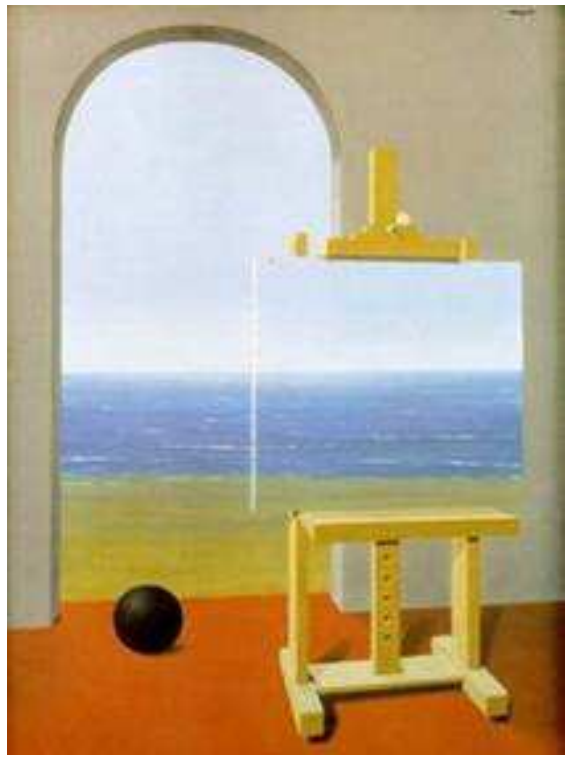

**Figure 4. *La Condition Humaine*, Magritte (1935).**

As was so well put forward by Bunge (2003), whereas Mechanism had proposed a unified picture of nature, the new world-view looks like a cubist painting. It is not only mosaic: it is also highly *counter-intuitive*.

*Virtual and hyper-real*

In common use, the word *virtual* often designates the absence of existence, a fake or illusory world, opposed to the *real*, material, concrete world. The term *virtual* comes from Medieval Latin *virtuälis*, equivalent to Latin *virtus*, which means virtue, force, power. For Lévy (1998) and for the Scholastics, virtual does not oppose to *real* but to the *actual* and is something that is in a potential state, not yet expressed or actualized.

In our so-called post-modern era, more modern technologies – like film, television, or the computer screen – created "realistic" images that did not rely necessarily on anything actual, but rather, by tricking the eye, on realities that are simultaneously constructed subjectively by the senses of the observer. People are now inundated by flickering images – legitimate cultural and material heirs from *the camera obscura*, the magic lantern, and the stereoscope – that forms the present discourse on virtual reality and virtual environments (Jones, 2006).

However, with such a multitude of technological resources at our disposal, virtual does not seem to be enough. We live in a *more to come* consumer culture, according to Eco (1986). For a game to be funnier, more exhilarating, more absorbing, its designer will make it "better." Even if the virtual environment implements physical laws, game designers may not want to follow real world rules and allow, for example, players to drive faster, jump farther or bounce harder than normal (Havok, 2008).

Those simulations that, to be labeled as true must *look* true, have the appearance of being more real that the original – much like Caesar's wife, who must not only *be* virtuous, but must *be seen* to be virtuous (Plutarch, 1919) – Eco (1986, p. 13) designated as *hyper-real*. Notice, however, that this is distinct from Baudrillard's (1983) hyperreality, which is a sense of reality created by technology and made so similar to true reality, as to be indistinguishable from it – not more real than reality, as in Eco's one. It should also not to be confused with *Hyperrealism*, a genre of painting and sculpture so meticulously detailed as to resemble a high resolution photograph.

*SL physics*

As discussed in the previous section, while SL implements various real world features such as wind, air resistance, terminal velocity, and settling of moving objects, SL physics is not a mere real world physics virtualization, as other important features such as water resistance were not implemented. On the other hand, various physical quantities have quite different definitions in SL when compared to the Newtonian physics ones, as seen in the previous section, and therefore, it is not either a mere implementation of idealized Newtonian physics as offered by the Havok engine.

Havok, while embodying Newtonian physical laws, also allows and even encourages the building of hyper-real simulations, a path apparently followed by SL implementation. As a matter of fact, Philip Rosedale, in an interview quoted in Eliëns et al. (2007), affirms that one of the main reasons for the success of SL is "the fact that it offers a set of capabilities, which are in many different ways superior to the real world." Therefore, SL deliberately chose to implement physics in a basically hyper-real way.

The author of the present study wants to stress, however, that SL can go far beyond that. The possibility of making an object non-physical, switching off the standard SL physics and programming new unusual physical behaviors through LSL scripts opens the door to imagination and experimentation. As two very pale examples, Havok (2008) suggests alternative gravity settings along an alternative axis. The Guide also suggests adding up to three times the normal gravity to a car with the basic effect of "accelerating the subjective time" (Havok, p. 330). One could use SL as a lab world, where one could easily make objects that repel themselves from one another or that follow strange dynamical laws. The experimenter would, however, not do it just for fun but mainly to liberate imagination and be able to critically question the closed mechanistic, positivistic physics taught in schools. These motivations, resembling those of the Surrealism approximation to modern physics, as discussed above, is the reason why the author prefers to use the adjective *surreal* when referring to the possible use of SL as a simulator.

To sum things up, Havok physics can be interpreted as hyper-real as well as the usual resident experience in SL. Yet, at the same time SL offers resources to build a surreal simulator in this metaverse. Its pedagogical implications are interesting as will be discussed in the next section.

## Pedagogical Implications

*Trinity: "The Matrix isn't real."*
*Cypher: "I disagree, Trinity. I think the Matrix can be more real than this world."* (Irwin, 2002).

Most learners – including many science majors – have difficulty in understanding physics concepts and models, both at the qualitative level and the quantitative formulation (Reif & Larkin, 1992). That difficulty often arises from alternative conceptions (Driver, 1989) built from their common personal experiences, based on a lifetime's experience and is very difficult to remediate with instructionist pedagogical strategies (Dede, 1995).

Papert, displeased with this inefficient learning process, has strongly defended the introduction of IT in the classroom as a means to actively engage students in constructing mental models and theories of the world since the 1970s. For Morgan & Morrison (1999), models are mediators between theory and the real world, between classroom abstract scientific knowledge and the student's concrete, empirical experience. It must be remembered that simulation is not a new concept, as scientists and programmers have used computers to simulate complex situations like rocket trajectories (ballistic motion), liquid flows (fluid dynamics) and other complicated projects probably since they started dropping off the assembly line (Havok, 2008, p. 371).

Papert, as early as 1980, offered a "Piagetian learning path into Newtonian laws of motion" (Papert, 1993, p. 123). For Papert, the phrase *laws of motion* usually raises difficult questions like What other laws of motion are there besides Newton's? (p. 124). To him, learners should be acquainted with other laws of motion, not so subtle and counter-intuitive. This would be viable in a physics microworld, where they could build an infinite variety of laws of motion, progressing, thus, from the historically and psychologically important Aristotelian ones to the "correct" Newtonian ideas and even to the more complex Einsteinian ones, via as many intermediate worlds as they wish (p. 125), in a way that short-minded teachers "may refuse to recognize as physics" (p. 122). And so many can be these worlds that "the logical distinction between the 'real world' and 'possible worlds' has been undermined" (Eco, 1986, p. 14), in the sense of immersion and experimentation, of distinction between *possibility* and *necessity* (Piaget, 1987). It is worth remembering that this experimental progression from Aristotelian physics to Newton mechanics closely resembles the psychogenetic succession investigated by Piaget & Garcia in their paramount work *Psychogenesis and the History of Science* (1989).

Various microworlds have been built, from the original Papert's Logo Turtles (1993) to present educational environments with *Augmented Reality* (Azuma et al., 2001). However, almost thirty years after Papert's proposal, besides the primitive diSessa's (Abelson, H. & diSessa, A., 1981) *Dinatarts*, to our knowledge there is no microworld implementation which allows the experimentation with physical laws, as conceived by Papert.

On the other hand, present high-performance computing and communications capabilities create a new possibility (Dede, 1995). It allows learners to immerse themselves in virtual, synthetic environments, like Alice walking through the looking glass, becoming *avatars* that can collaborate and learn-by-doing using virtual artifacts to construct knowledge (Walker, 1990). This possibility shifts the focus of constructivism, "'magically' shaping the fundamental nature of how learners experience their physical and social context" (Dede, p. 1). According to Dede, "this instructional approach enhances students' ability to apply abstract knowledge by situating education in authentic, virtual contexts similar to the environments in which learners' skills will be used" in the future (p. 1).

This mediation may be even improved in an immersive 3-D metaverse as SL, shifting the education from the traditional classroom layout and dynamics to a multisensory learning environment "where students can be part of the system that is being studied" (Calogne & Hiles, 2007). In fact, "Havok provides low-level access to core functionality so that you can construct complex physical behaviors that are specific to your game and don't come as standard with Havok" (Havok, 2008, p. 96).

By comparing SL with a traditional simulation environment like Modellus, the SL physics hyper-reality could, as previously discussed, constitute at first sight an obstacle to its utilization as a simulation environment for physics teaching. However, it is this author's understanding that this same hyper-reality and, even better, its surreal potential, is a golden opportunity. An SL physics lab could allow a surreal experimentation with successive or generalized physical laws, as proposed by Papert, in a Piagetian historical-psychogenetic framework (Piaget & Garcia, 1989) which could be pedagogically effective. At the same time it would allow extremely rich epistemological discussions on the nature of physical concepts and on issues such as *what is a physical law?* (Feynman, 1967), *what is Science?*, *what kind of reality does Science describe?*, and *who makes the Science decisions?*, to name a few.

That is the proposal of the project this author is starting.

## Conclusions

*Trinity: "No one has ever done anything like this."*
*Neo: "That's why it's going to work."* (Irwin, 2002).

Garcia, discussing the new Internet "upload" concept for education, warns that:

"Some authors foresee that in few years' time textbooks will be totally replaced by electronic media. Some of them believe that even classrooms will disappear. We cannot predict what will happen in the future with our educational models, but changes will be enormous and many things that we do today will belong in museums. We have to be prepared"(2008).

In this work it was shown that, after all, SL physics is neither the Galilean/Newtonian idealized physics nor a real world physics virtualization; rather, it concludes that SL physics is *hyper-real* and provides resources for building a *surreal* physics lab that allows experimentation with successive or generalized physical laws. Also, it still provides a rich environment for classroom epistemological discussions around the reality/unreality of the physical laws seen in school, in good accordance with Papert's (1993) never-implemented proposal.

That was the reason for our travel in Second Life hyperreality, "in search of instances [. . .] where the boundaries between game and illusion are blurred .[ . .] and falsehood is enjoyed in a situation of 'fullness'" (Eco, 1986, p. 8).

**Acknowledgements**

The author of the present study deeply acknowledges the enlightening comments from Prof. Dr. Maurício Rosa (Ulbra/PPGECIM) and the detailed and constructive comments from two anonymous reviewers of my ambitious manuscript which helped me to make my points clearer and stronger, as well as Felix Nonnenmacher's meticulous copyediting that made a nicely readable text out of my awkward and confusing phrases.